\documentclass[prd,twocolumn,showpacs,preprintnumbers,amsmath,amssymb,superscriptaddress,floatfix,nofootinbib]{revtex4}

\usepackage{graphicx}
\usepackage{amsmath}
\usepackage{amsfonts}
\usepackage{amssymb}
\usepackage{color}
\usepackage{multirow}
\usepackage[colorlinks, citecolor=blue,anchorcolor=red,menucolor=red, linkcolor=red,filecolor=red,runcolor=red,urlcolor=blue,frenchlinks=red]{hyperref}

\begin{document}

\title{Strong decays of the $X(2500)$ newly observed by the BESIII Collaboration}

\author{Ting-Ting Pan}
\affiliation{Department of Physics, Zhengzhou University, Zhengzhou, Henan 450001, China}

\author{Qi-Fang L\"{u}} \email{lvqifang@ihep.ac.cn}
\affiliation{Institute of High Energy Physics, Chinese Academy of Sciences, Beijing 100049, China}

\author{En Wang} \email{wangen@zzu.edu.cn}
\affiliation{Department of Physics, Zhengzhou University, Zhengzhou, Henan 450001, China}

\author{De-Min Li} \email{lidm@zzu.edu.cn}
\affiliation{Department of Physics, Zhengzhou University, Zhengzhou, Henan 450001, China}

\begin{abstract}
Basing on the observation of a new $J^{PC}=0^{-+}$ state, $X(2500)$, in the partial wave analysis of the decay $J/\psi\rightarrow\gamma X\rightarrow \gamma\phi\phi$ performed by the BESIII Collaboration, we have evaluated the strong decays of the $X(2500)$ as the $4^1S_0$ and $5^1S_0$ $s\bar{s}$ states in the $^3P_0$ model of meson decay. The predicted total decay width for the $4^1S_0$ $s\bar{s}$ is about 894.5 MeV, and the one for the $5^1S_0$ $s\bar{s}$ is about 271.1 MeV, which is in agreement with the experimental data $\Gamma_{X(2500)}=230^{+64+56}_{-35-33}$ MeV. By considering the mass and the total strong decay width of the $X(2500)$, we propose that the $X(2500)$ state can be interpreted as a candidate of the $5^1S_0$ $s\bar{s}$ state.

\end{abstract}

\maketitle

\section{Introduction}

Recently, the BESIII Collaboration has preformed a partial wave analysis of the decay $J/\psi\rightarrow\gamma X\rightarrow \gamma\phi\phi$ to study the intermediate states~\cite{Ablikim:2016hlu}. Besides the confirmation of the $\eta(2225)$, two additional pseudoscalar states, $\eta(2100)$ and $X(2500)$, are also reported.
The $\eta(2100)$ has been listed in Particle Data Group book as the further state~\cite{Agashe:2014kda}, which was found in the $J/\psi \to 4\pi \gamma$ process~\cite{Bisello:1988as}. The $X(2500)$ is the newly observed state with the significance of $8.8~\sigma$, and the mass and decay width are
\begin{eqnarray}
 &&M_{X(2500)}=2470^{+15+101}_{-19-23}~{\rm MeV}, \nonumber \\  &&\Gamma_{X(2500)}=230^{+64+56}_{-35-33}~{\rm MeV}.
\end{eqnarray}

In the light pseudoscalar sector, the $1^1S_0$ meson nonet ($\pi, \eta, \eta^\prime,$ and $K$) as well as the $2^1S_0$ members [$\pi(1300)$, $\eta(1295)$, $\eta(1475)$, and $K(1460)$] have been well established~\cite{Agashe:2014kda}. In the Refs.~\cite{Li:2008mza,Liu:2010tr,Yu:2011ta}, the $\pi(1800)$ and $K(1830)$, together with the $X(1835)$ and $\eta(1760)$ observed by the BES Collaboration~\cite{Ablikim:2005um, Ablikim:2006ca}, are suggested to constitute the $3^1S_0$ meson nonet. In addition, the $\pi(2070)$, $\eta(2100)$ and  $\eta(2225)$ are interpreted as the members of the $4^1S_0$ meson nonet in Refs.~\cite{Li:2008we,Li:2008et,Yu:2011ta}.
The $X(2370)$ observed in $J/\psi \to \gamma \pi^+ \pi^- \eta$~\cite{Ablikim:2010au} was suggested to be a good isoscalar  candidate of the $5^1S_0$ nonet~\cite{Yu:2011ta}, and  the $\pi(2360)$ observed in a partial wave analysis of $p\bar{p}\to \eta\eta\pi$ process~\cite{Anisovich:2001pp} was interpreted to be the isovector candidate of the $5^1S_0$ nonet~\cite{Anisovich:2005dt}.

\begin{figure}[htpb]
\centering
\includegraphics[scale=0.2]{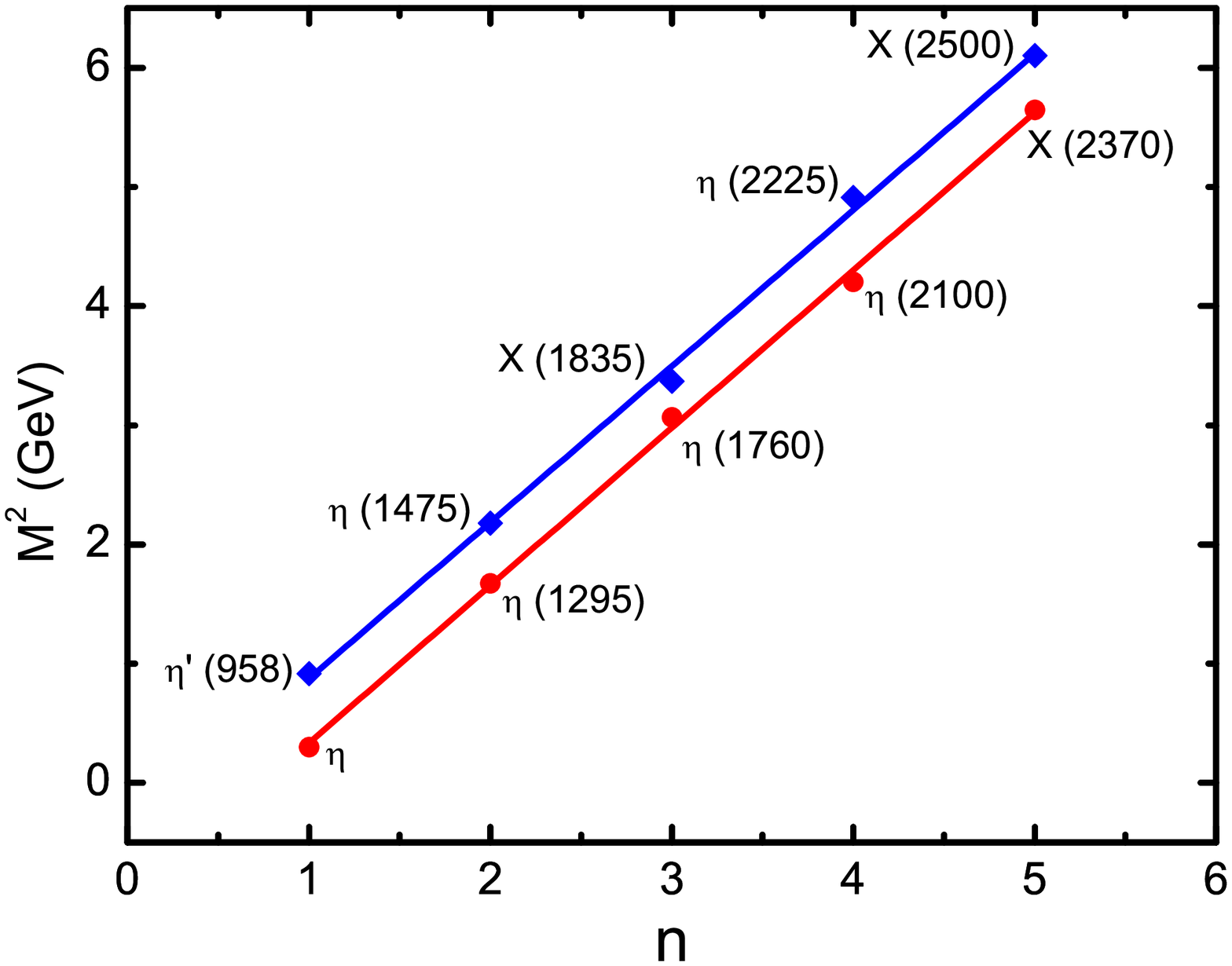}
\includegraphics[scale=0.2]{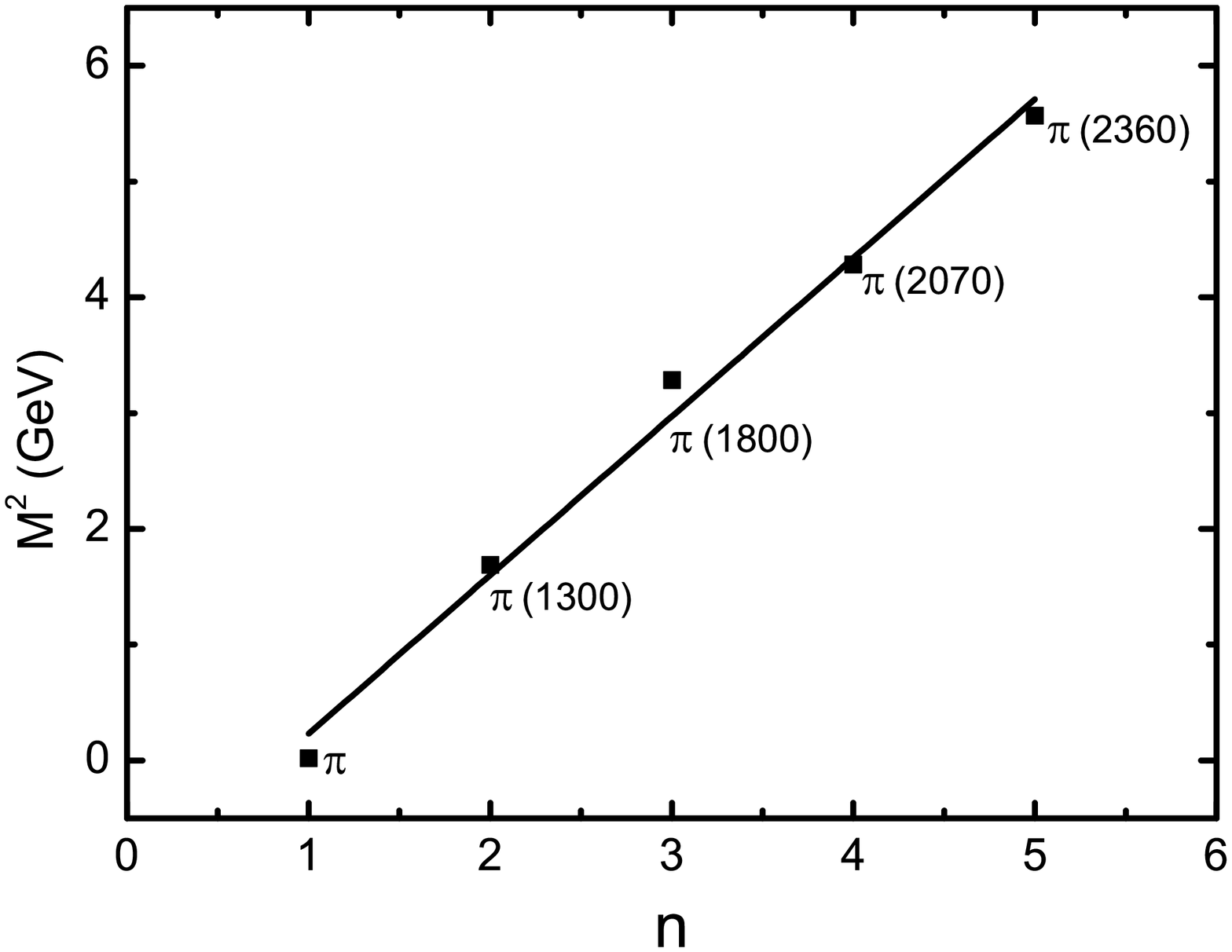}
\vspace{0.1cm}\caption{The Regge trajectories for the $n^1S_0$ meson
mass spectrum with $M^2=M^2_0+(n-1)\mu^2$ $(\mu^2=1.31$, $1.32$, 1.36 GeV$^2$ for the $\eta^\prime$, $\eta$, and $\pi$-trajectories, respectively.
The masses of $\eta(2100)$, $\eta(2225)$, and $X(2500)$ are from the BESIII results~\cite{Ablikim:2016hlu},
the $X(2370)$ mass is from the BESIII results~\cite{Ablikim:2010au}. All the other states masses are taken from
PDG~\cite{Agashe:2014kda} }
 \label{MM2}

\end{figure}

It is suggested that the light mesons
could be grouped into the following Regge trajectories\cite{mesontrajectory}
\begin{eqnarray}
M^2_n=M^2_0+(n-1)\mu^2,
\label{trajectory}
\end{eqnarray}
where $M_0$ is the lowest-lying meson mass, $n$ is the radial
quantum number, and $\mu^2$ is the slope parameter of the
corresponding trajectory. In Fig.~\ref{MM2}, we plot the $0^{-+}$
trajectory on the plane of $(n,M^2)$ adopting the
relation of  Eq.~(\ref{trajectory}). It shows that the $\eta$, $\eta(1295)$,
$\eta(1760)$, $\eta(2100)$, and $X(2370)$ ($\pi$, $\pi(1300)$,
$\pi(1800)$,  $\pi(2070)$, and  $\pi(2360)$) can be well
accommodated into a trajectory of the isoscalar (isovector) states,
and the $\eta'$, $\eta(1475)$, $X(1835)$, $\eta(2225)$, and
$X(2500)$ approximately populate a common trajectory, which suggests
that, in the presence of the $X(2370)$ being the $5^1S_0$ isoscalar
state~\cite{Yu:2011ta}, the $X(2500)$ could be another $5^1S_0$
isoscalar state.  If one accepts that the $\pi(2360)$, $X(2370)$, and $X(2500)$ belong to the $5^1S_0$ meson nonet, the nearly degenerate masses of the
$X(2370)$ and the $\pi(2360)$ would imply that the $X(2370)$ is mainly $(u\bar{u}-d\bar{d})/\sqrt{2}$. No observation of
the $X(2370)$ state in the $J/\psi\rightarrow \gamma\phi\phi$
process~\cite{Ablikim:2016hlu} favors this argument. Therefore, as the orthogonal partner of the $X(2370)$,
the $X(2500)$ could be treated as the $5^1S_0$ $s\bar{s}$ state based on its mass~\cite{Masjuan:2012gc}.

The mass information alone is insufficient to identify the $X(2500)$ as the pseduoscalar meson excitation, because the mass of the lowest pseudoscalar glueball predicted by the
Lattice QCD is in the range of 2.3$\sim$2.6
GeV~\cite{Chen:2005mg,Bali:1993fb,Morningstar:1999rf}, which is also
in consistent with the $X(2500)$ mass. We shall discuss the possibility of the $X(2500)$ being the ordinary $5^1S_0$ $s\bar{s}$ by studying its strong decay properties. It is natural and
necessary to exhaust the possible $q\bar q$ descriptions of a newly observed state before restoring to the more exotic assignments.

In this work we study the strong decays of the $X(2500)$ in the
$^3P_0$ model of meson decay, assuming it being
the $5^1S_0$ $s\bar{s}$ state. The $4^1S_0$ $s\bar{s}$ assignment of the $X(2500)$ is
also discussed. We calculate the partial decay widths and
total decay width by taking into account 21 decay channels, and discuss
the dependence of predictions on the $X(2500)$ mass.
Our result of the total decay width of $X(2500)$ indicates that the $X(2500)$ can be regarded as the candidate of
the $5^1S_0$ $s\bar{s}$ state.

This paper is organized as follows. In Sec.~\ref{sec:formalism},
we will present a brief review of the $^3P_0$ model of meson decay. The
results and the discussions of the strong decays of the $X(2500)$
state are shown in Sec.~\ref{sec:result}. Finally, the summary
is given in Sec.~\ref{sec:summary}.

\section{THE $^3P_0$ MODEL OF MESON DECAY }
\label{sec:formalism}

In this section, we will give a brief introduction of the $^3P_0$ model.
The $^3P_0$ model, also known as the the quark-pair creation model (QPC), was originally introduced by Micu~\cite{Micu:1968mk} and further developed by Le Yaouanc~$et$ $al.$\cite{LeYaouanc:1972vsx,LeYaouanc:1973ldf,LeYaouanc:1977fsz,LeYaouanc:1977gm,LeYaouanc:alo}, and has been widely applied to study hadron strong decays with considerable success~\cite{W.Roberts:few,Blundell:1996as,a1,a2,a3,a4,a5,a6,a7,a8,a9,a10,a11,a12,a13,a14,a15,a16,a17,a18,Li:2009rka,Lu:2014zua}.

\begin{figure}[htb]
\centering
\includegraphics[scale=0.25]{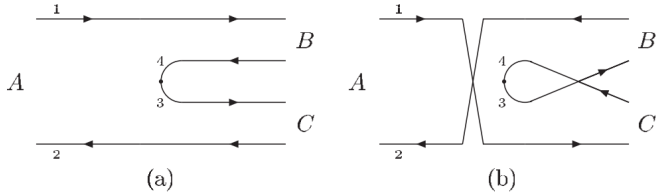}
\caption{The two possible diagrams contributing to $A\rightarrow BC$ in the $^3P_0$ modelz: (a) the quark within the meson A combines with the created antiquark to form the meson B, the antiquark within in the meson A combines with the created quark to form the meson C; (b) the quark within the meson A combines with the created antiquark to form the meson C, the antiquark within the meson A combines with the created quark to form the meson B.}
\label{fig:3p0}
\end{figure}

In the $^3P_0$ model, the strong decays occur by producing a quark-antiquqrk pair with the vacuum quantum number $J^{PC}=0^{++}$. The newly produced quark-antiquark pair, together with the $q\bar{q}$ within the initial hadron, regroups into two outgoing hadrons in all possible quark rearrangement ways. The decays process for the meson case can be depicted in Fig.~\ref{fig:3p0}

For the meson decay process,
\begin{equation}
A(P_A) \rightarrow B(P_B) + C(P_C),
\end{equation}
the transition operator $T$ can be written by,
\begin{eqnarray}
T=-3\gamma\sum_m\langle 1m\,1-m|00\rangle\int
{\rm d}^3\vec{p}_3\,{\rm d}^3\vec{p}_4\,\delta^3(\vec{p}_3+\vec{p}_4)\nonumber\\
{\cal{Y}}^m_1\left(\frac{\vec{p}_3-\vec{p}_4}{2}\right
)\chi^{34}_{1-m}\phi^{34}_0\omega^{34}_0b^\dagger_3(\vec{p}_3)d^\dagger_4(\vec{p}_4),
\end{eqnarray}
where the dimensionless parameter $\gamma$ represents the probability of the quark-antiquark pair with $J^{PC}=0^{++}$ creation from the vacuum,  $\vec{p}_3$ ($\vec{p}_4$) is the momentum of the created quark (antiquari) $q_3$ ($q_4$), and $\phi^{34}_0$, $\omega^{34}_0$, and $\chi^{34}_{1-m}$ are the flavor, color, and spin wave functions of the $q_3\bar{q_4}$ pair, respectively.
The solid harmonic polynomial ${\cal{Y}}^m_1(\vec{p})\equiv|p|^1Y^m_1(\theta_p, \phi_p)$ reflects the momentum-space distribution of the $q_3\bar{q_4}$ pair.

The helicity amplitude ${\cal{M}}^{M_{J_A}M_{J_B}M_{J_C}}
(\vec{P})$ is defined as,
\begin{eqnarray}
\langle
BC|T|A\rangle=\delta^3(\vec{P}_A-\vec{P}_B-\vec{P}_C){\cal{M}}^{M_{J_A}M_{J_B}M_{J_C}}(\vec{P}),\end{eqnarray}
where $\vec{P}_A$, $\vec{P}_B$, and $\vec{P}_C$ are the 3-momentum of the mesons $A$, $B$, and $C$, respectively. The $|A\rangle$, $|B\rangle$, and $|C\rangle$ denote the mock meson states, and the mock meson $|A\rangle$ is defined by~\cite{Hayne:1981zy}
\begin{eqnarray}
&&\left. | A(n^{2S_A+1}_A L_{A} J_A M_{J_A})(\vec{P}_A) \right\rangle
 \nonumber\\
 &\equiv & \sqrt{2E_A}\sum_{M_{L_A},M_{S_A}} \left\langle L_A M_{L_A} S_A M_{S_A}|J_A M_{J_A} \right\rangle  \nonumber\\
& \times&  \int {\rm d}^3\vec{p}_A\psi_{n_A L_AM_{L_A}}(\vec{p}_A)\chi^{12}_{S_AM_{S_A}}
\phi^{12}_A\omega^{12}_A \nonumber \\
&\times&  \left| q_1 \left( \frac{m_1}{m_1+m_2}\vec{P}_A+\vec{p}_A \right)\bar{q}_2 \left({\frac{m_2}{m_1+m_2}}\vec{P}_A-\vec{p}_A \right) \right\rangle, \nonumber \\
\end{eqnarray}
where $m_1$ and $m_2$ ($\vec{p}_1$ and $\vec{p}_2$) are the masses (momenta) of the quark $q_1$ and the antiquark $\bar{q}_2$, respectively; $\vec{P}_A=\vec{p}_1+\vec{p}_2$, $\vec{p}_A=(m_2\vec{p}_1-m_1\vec{p}_2)/(m_1+m_2)$;
$\chi^{12}_{S_AM_{S_A}}$, $\phi^{12}_A$, $\omega^{12}_A$, and
$\psi_{n_AL_AM_{L_A}}(\vec{p}_A)$ are the spin, flavor, color, and
space wave functions of the meson $A$ composed of $q_1\bar{q}_2$ with total energy $E_A$, respectively. $n_A$ is the radial quantum number of the meson $A$. $S_A=s_{q_1}+s_{\bar{q_2}}$, $J_A=L_A+S_A$, $s_{q_1}(s_{\bar{q_2}})$ is the spin of $q_1(\bar{q_2})$, and $L_A$ is the relative orbital angular momentum between $q_1$ and $\bar{q_2}$. $\langle L_A M_{L_A} S_AM_{S_A}|J_AM_{J_A}\rangle$ is a Clebsch-Gordan coefficient. The mock meson satisfies the normalization condition,
\begin{eqnarray}
&&\langle A(n_A^{2S_A+1}L_A J_AM_{J_A})(\vec{p}_A)|A(n_A^{2S_A+1}L_AJ_AM_{J_A})(\vec{p}_A^{\,\prime})\rangle
\nonumber
\\&=& 2E_A\delta^3(\vec{p}_A-\vec{p}_A^{\,\prime}).
\end{eqnarray}

In the center of mass frame (c.m.) of meson $A$, the explicit form of the helicity amplitude can be written as,
\begin{eqnarray}
&&{\cal{M}}^{M_{J_A}M_{J_B}M_{J_C}}(\vec{P})=\gamma\sqrt{8E_A E_B E_C}\nonumber\\
&&\times \sum_{M_{L_A}}\sum_{M_{S_A}}\sum_{M_{L_B}}\sum_{M_{S_B}}\sum_{M_{L_C}}\sum_{M_{S_C}}\sum_{m}
\langle L_A M_{L_A} S_AM_{S_A}|J_AM_{J_A}\rangle\nonumber\\
&&\times\langle L_B M_{L_B} S_BM_{S_B}|J_BM_{J_B}\rangle\langle L_C M_{L_C} S_CM_{S_C}|J_CM_{J_C}\rangle\nonumber\\
&&\times\langle 1m1-m|00\rangle\langle \chi^{14}_{S_BM_{S_B}}\chi^{32}_{S_CM_{S_C}}|\chi^{12}_{S_AM_{S_A}}\chi^{34}_{1-m}\rangle\nonumber\\
&&\times[\boldsymbol{f}_1I(\vec{P},m_1,m_2,m_3)\nonumber\\
&&+(-1)^{1+S_A+S_B+S_C}\boldsymbol{f}_2 I(-\vec{P},m_2,m_1,m_3)],
\end{eqnarray}
where the two terms $\boldsymbol{f}_1=\langle \phi^{14}_B\phi^{32}_C|\phi^{12}_A\phi^{34}_0\rangle$ and $\boldsymbol{f}_2=\langle \phi^{32}_B\phi^{14}_C|\phi^{12}_A\phi^{34}_0\rangle$ correspond to the contributions from Fig.~\ref{fig:3p0}(a) and Fig.~\ref{fig:3p0}(b), respectively, and the momentum space integral is,
\begin{eqnarray}
&&I(\vec{P},m_1,m_2,m_3) \nonumber \\&=&\int {\rm d}^3\vec{p}\,\psi^*_{n_BL_BM_{L_B}}\left({
\frac{m_3}{m_1+m_2}}\vec{P}_B+\vec{p}\right)\nonumber\\&&\times\psi^*_{n_CL_CM_{L_C}}\left({
\frac{m_3}{m_2+m_3}}\vec{P}_B+\vec{p}\right)\nonumber\\&&\times\psi_{n_AL_AM_{L_A}}\left(\vec{P}_B+\vec{p}\right){\cal{Y}}^m_1(\vec{p}),
\end{eqnarray}
where $\vec{P}={\vec{P}}_B=-{\vec{P}}_C$, $\vec{p}=\vec{p}_3$, and $\psi$ is the meson wave function in momentum space.
The spin overlap in terms of $9j$ symbol can be given by
\begin{eqnarray}
&&\langle\chi^{14}_{S_BM_{S_B}}\chi^{32}_{S_CM_{S_C}}|\chi^{12}_{S_AM_{S_A}}\chi^{34}_{1-m}\rangle=\nonumber\\
&&\times\sum_{S,M_S}\langle S_B M_{S_B} S_CM_{S_C}|SM_S\rangle\nonumber\\
&&\times\langle S_A M_{S_A} 1-m|SM_S\rangle(-1)^{S_C+1}\nonumber\\
&&\times\sqrt{3(2S_A+1)(2S_B+1)(2S_C+1)}\nonumber \\ &&\times \left\{
                                          \begin{array}{ccc}
                                            1/2 & 1/2 & S_A \\
                                           1/2 & 1/2& 1 \\
                                           S_B &S_C & $S$ \\
                                          \end{array}
                                        \right\}
\end{eqnarray}

The partial wave amplitude ${\cal{M}}^{LS}(\vec{P})$ can be obtained from the helicity amplitude,
\begin{eqnarray}
{\cal{M}}^{LS}(\vec{P})&=&
\sum_{M_{J_B}}\sum_{M_{J_C}}\sum_{M_S}\sum_{M_L}
\langle LM_LSM_S|J_AM_{J_A}\rangle \nonumber\\
&&\langle
J_BM_{J_B}J_CM_{J_C}|SM_S\rangle\nonumber\\
&&\times\int
d\Omega\,\mathcal{Y}^\ast_{LM_L}{\cal{M}}^{M_{J_A}M_{J_B}M_{J_C}}
(\vec{P}), \label{pwave}
\end{eqnarray}

Because of  different choices of pair-production vertex, phase space convention, employed meson wave function, various $^3P_0$ models exist in literatures. In this article, we restrict to the simplest vertex as introduced originally by Micu~\cite{Micu:1968mk} which assumes a spatially constant pair-production strength $\gamma$, adopt the relativistic phase space, and employ the simple harmonic oscillator (SHO) approximation for the meson space wave functions which are commonly used in evaluating
the light mesons strong decays~\cite{Li:2008mza,Yu:2011ta,Li:2008we,Li:2008et,a7,a8,a9}.
With the relativistic phase space, the decay width
$\Gamma(A\rightarrow BC)$ can be expressed as follows,
\begin{eqnarray}
\Gamma(A\rightarrow BC)= \frac{\pi
|\vec{P}|}{4M^2_A}\sum_{L,S}\left|{\cal{M}}^{LS}(\vec{P})\right|^2, \label{width1}
\end{eqnarray}
where $M_A$, $M_B$, and $M_C$ are the masses of the meson $A$, $B$,
and $C$, respectively, and
\begin{equation}
|\vec{P}|=\frac{ \sqrt{[M^2_A-(M_B+M_C)^2][M^2_A-(M_B-M_C)^2]}}{2M_A}.
\end{equation}

Under the SHO approximation, the meson space wave function is
\begin{eqnarray}
\psi_{nLM_L}(\boldsymbol{p})=R_{nL}^{\rm SHO}(p)\mathcal{Y}_{LM_L}(\Omega_p),
\end{eqnarray}
where the radial wave function is given by
\begin{eqnarray}
R_{nL}^{\rm SHO}(p)=&&\frac{(-1)^n(-i)^L}{\beta^{3/2}}\sqrt{\frac{2n!}{\Gamma(n+L+3/2)}}\nonumber\\
&&\times\left(\frac{p}{\beta}\right)^Le^{-(p^2/2{\beta}^2)}L_n^{L+(1/2)}\left(\frac{p^2}{\beta^2}\right).
\end{eqnarray}
Here $\beta$ is the SHO wave function scale parameter, and $L_n^{L+(1/2)}\left({p^2}/{\beta^2}\right)$ is an associated Laguerre polynomial.

\section{DECAYS OF THE $4^1S_0$ AND $5^1S_0$ $s\bar{s}$ STATES IN THE $^3P_0$ MODEL}
\label{sec:result}
\begin{table}
\begin{center}
\caption{ \label{tab:decay}Decay widths of the $5^1S_0$ and $4^1S_0$ $s\bar{s}$ states in the $^3P_0$ model (in MeV). The initial state mass is set to 2470 MeV.}
\begin{tabular}{c|c|ccc}
\hline\hline
 Channel      & $i$           & Mode            & $\Gamma_i(4^1S_0)$     & $\Gamma_i(5^1S_0)$ \\
\hline
 $0^-\rightarrow 0^-0^+$                 & ch1    & $\eta f_0(980)$         & 4.03                 & 0.04  \\
       & ch2 & $\eta^\prime f_0(980)$     & 7.31               &1.62   \\
  $ $                      & ch3  & $\eta(1475) f_0(980)$      & 13.46               &18.85 \\
  $ $                        & ch4  & $\eta f_0(1710)$          & 2.89               & 1.23 \\
  \hline
  $0^-\rightarrow 1^-1^- $   & ch5  &$\phi(1020) \phi(1020)$     &2.46               &0.01\\
  \hline
  $0^-\rightarrow 0^-2^+ $    & ch6  &$\eta f_2^\prime(1525)$    &57.84               &9.61\\
  \hline
  $0^-\rightarrow 0^-0^+ $     & ch7  &$K K_0^*(1430)$           &11.00                &1.17\\
  $ $                          & ch8   &$K K_0^*(1950)$          &31.78                &22.22\\
  \hline
  $0^-\rightarrow 1^-1^+ $      & ch9   &$K^* K_1(1270)$          &35.42             &8.76\\
  $ $                           & ch10  &$K^*K_1(1400)$          &86.66               &18.13\\
  \hline
  $ $                          & ch11  &$KK^*$                    &9.13              &0.06\\
  $ $                          & ch12  &$K(1460) K^*$              &119.20             &29.43\\
  $0^-\rightarrow 0^-1^- $      & ch13 &$K K^*(1410)$              &7.80            &13.38\\
  $ $                           & ch14  &$K K^*(1680)$              &6.93           &2.29\\
  $ $                           & ch15 &$K  K(1830)$                &116.98          &72.21\\
  \hline
   $0^-\rightarrow 1^-1^- $     & ch16 &$K^* K^*$                 &15.68          &1.37\\
   & ch17 &$K^* K^*(1410)$          &223.77        &42.44\\
   \hline
    $0^-\rightarrow 0^-2^+ $     & ch18 &$K K^*_2(1430)$          &37.30         &0.43\\
    $ $                        & ch19   &$K K^*_2(1980)$          &0.01        &0.01\\
  \hline
  $0^-\rightarrow 1^-2^+ $        & ch20 &$K^* K^*_2(1430)$       &83.36          &18.59\\
  \hline
  $0^-\rightarrow 0^-3^- $       & ch21  &$KK^*_3(1780)$         &21.44         &9.25\\
  \hline
 \multicolumn{3}{c|}{Total width}     &    894.45    & 271.10\\
  \hline
  \multicolumn{3}{c|}{BESIII data}  & \multicolumn{2}{c}{$230^{+64+56}_{-35-33} $ }\\
\hline\hline
\end{tabular}
\end{center}
\end{table}

\begin{figure}[htpb]
\includegraphics[scale=0.7]{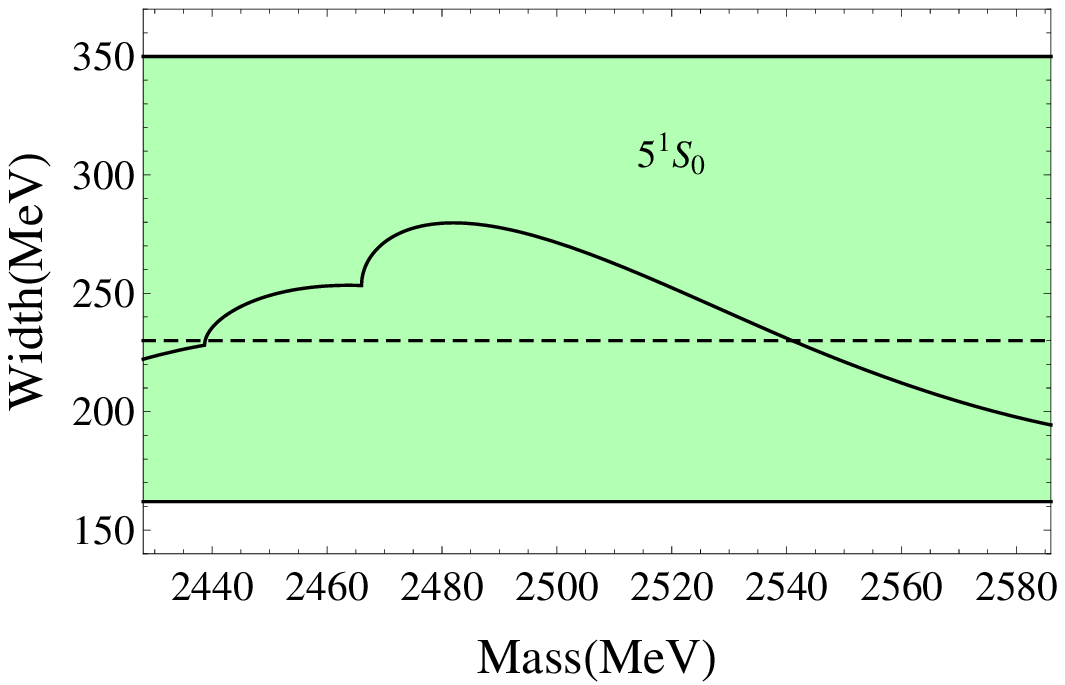}
\includegraphics[scale=0.7]{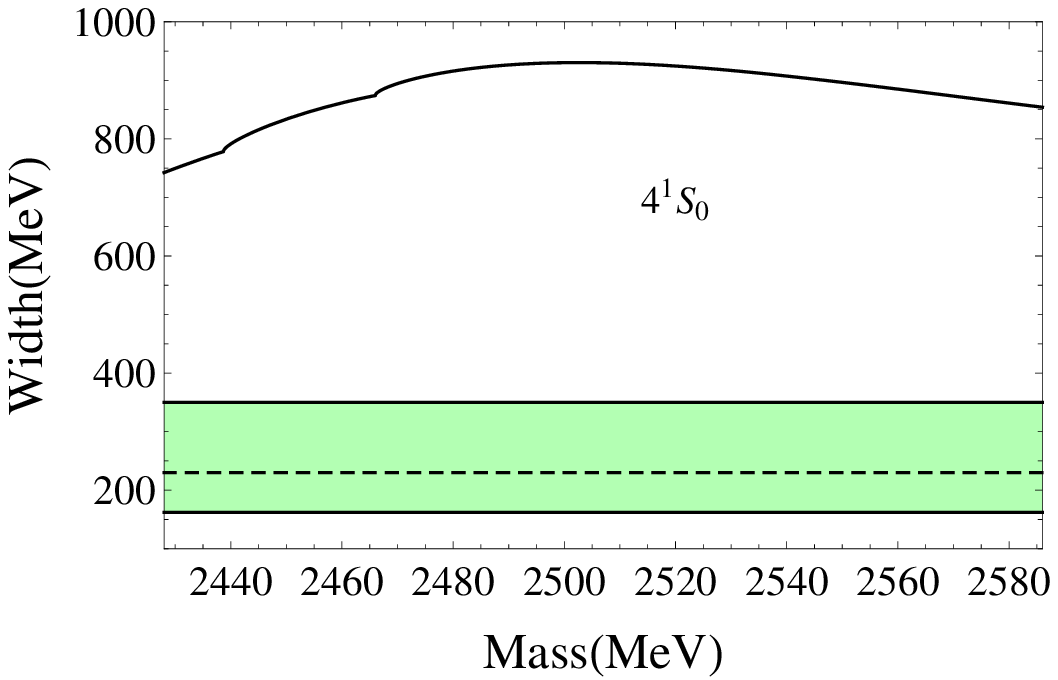}
\vspace{0.0cm}\caption{The dependence of the total widths of the $5^1S_0$ and $4^1S_0$ $s\bar{s}$ states on the initial state mass in the $^3P_0$ decay model. The dashed line with a green band denotes the BESIII data.}\label{fig:twidth}
\end{figure}

\begin{figure*}[htpb]
\includegraphics[scale=0.6]{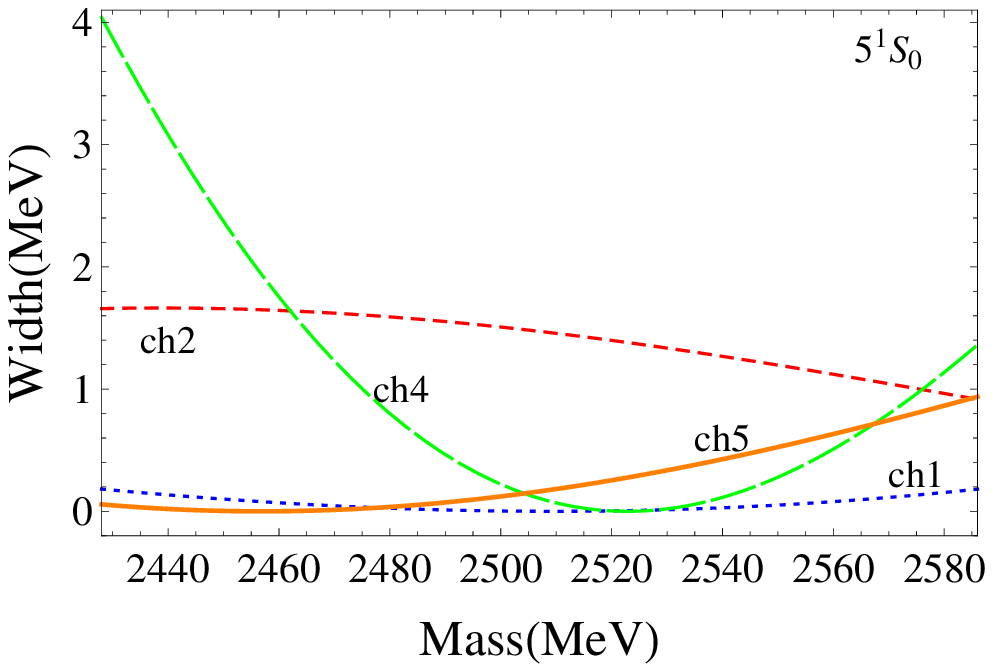}
\includegraphics[scale=0.6]{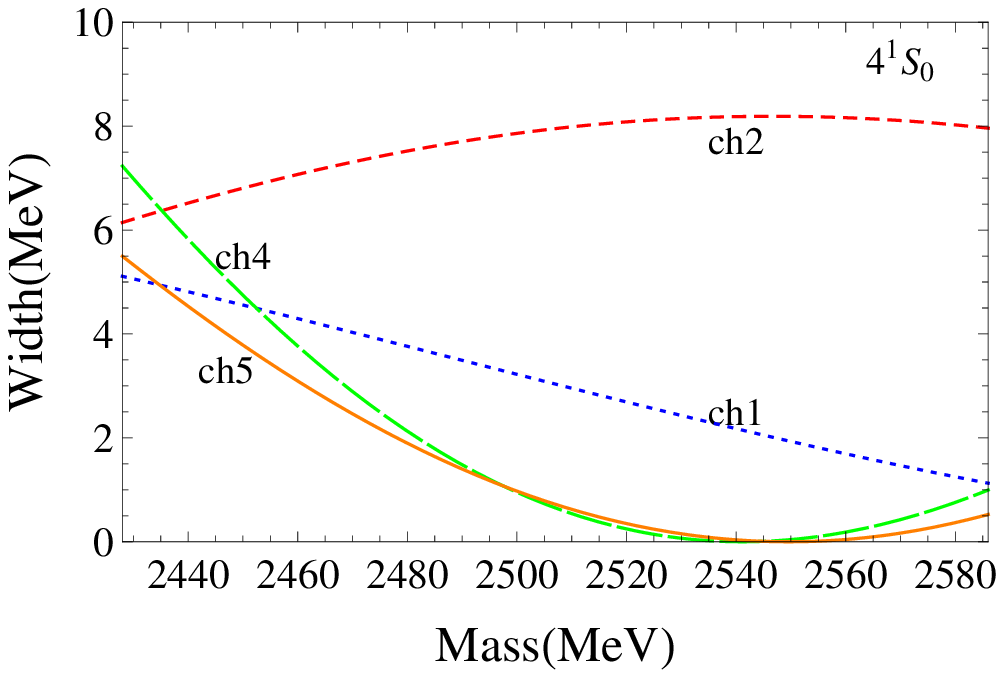}
\includegraphics[scale=0.6]{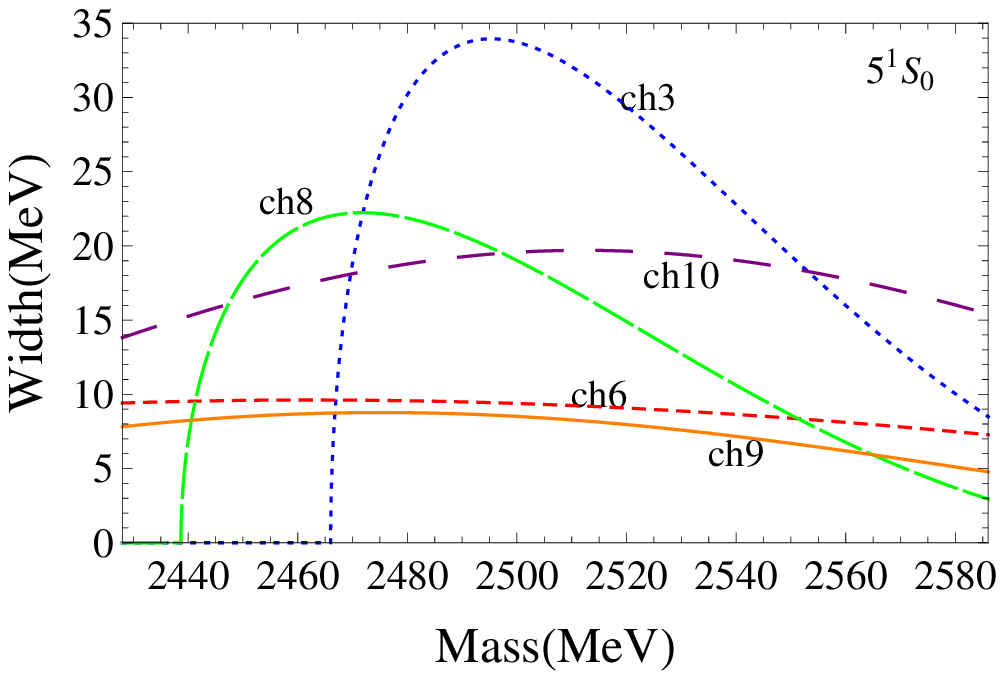}
\includegraphics[scale=0.6]{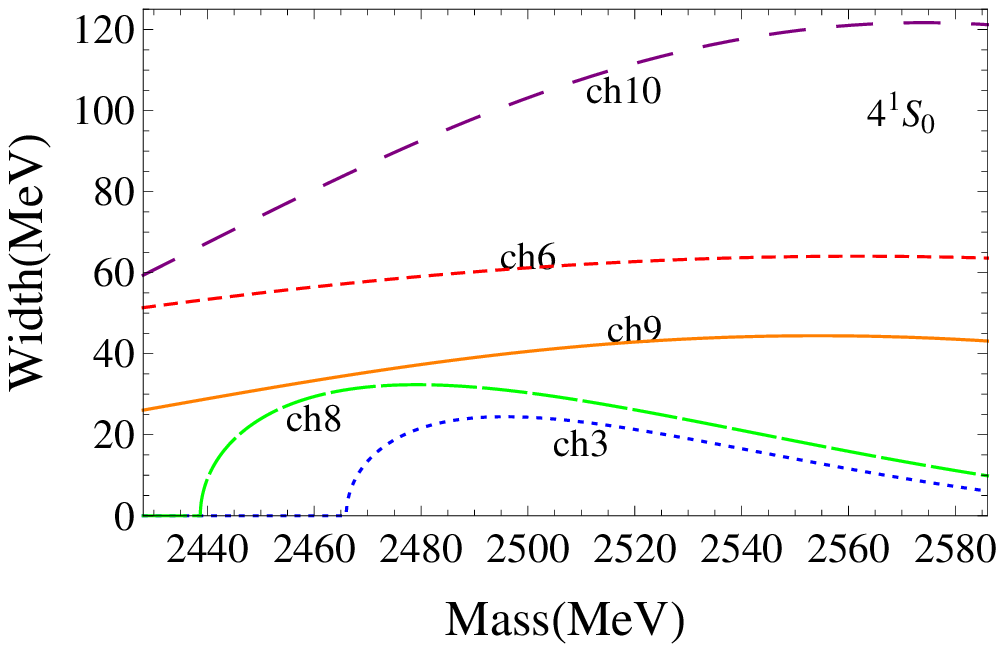}
\includegraphics[scale=0.6]{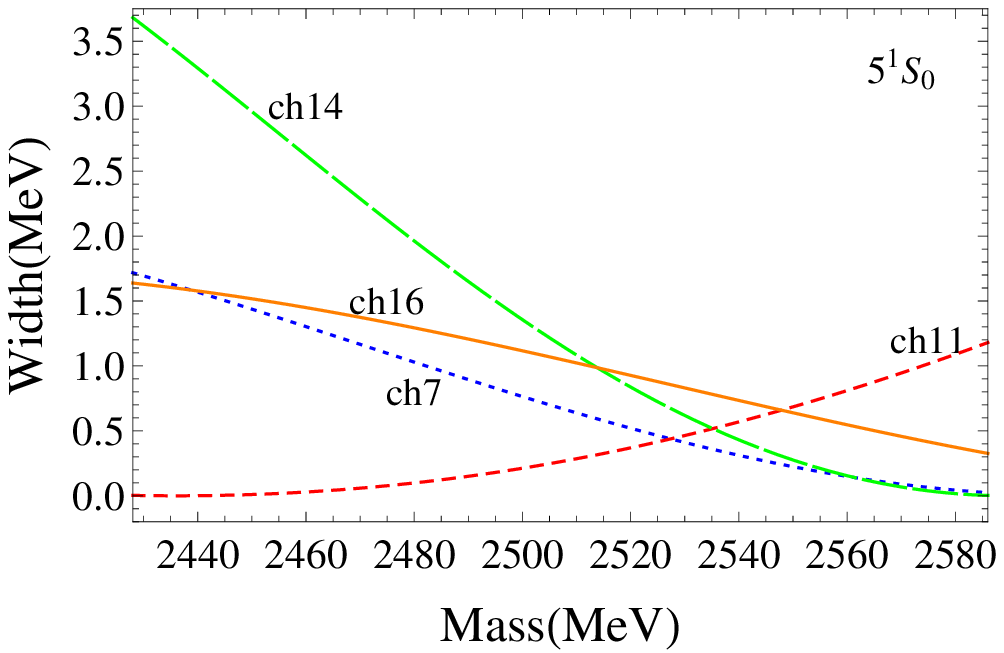}
\includegraphics[scale=0.6]{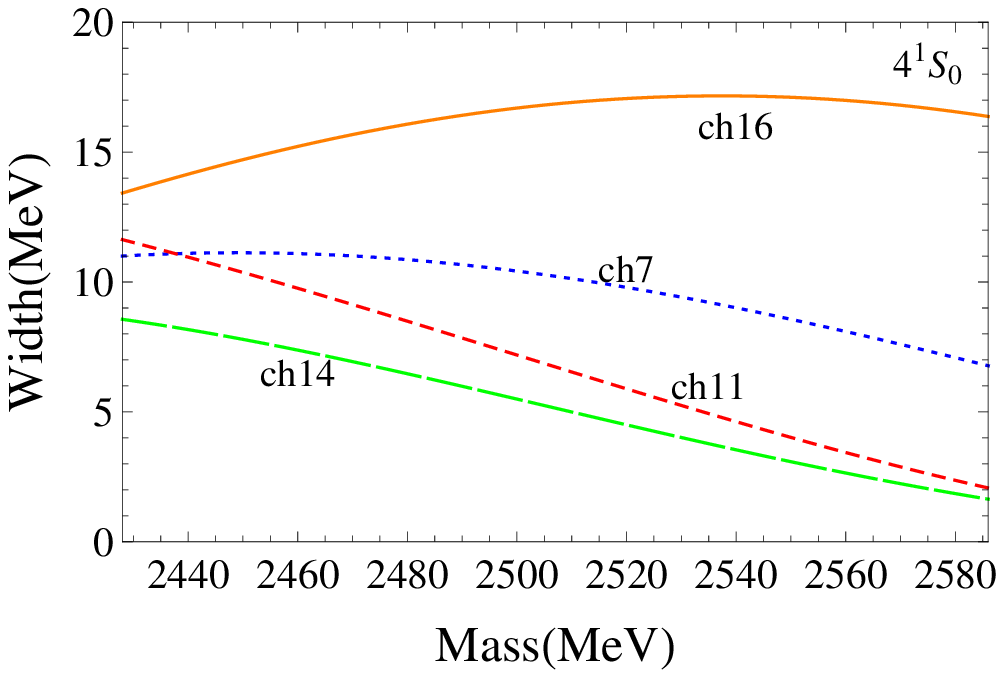}
\includegraphics[scale=0.6]{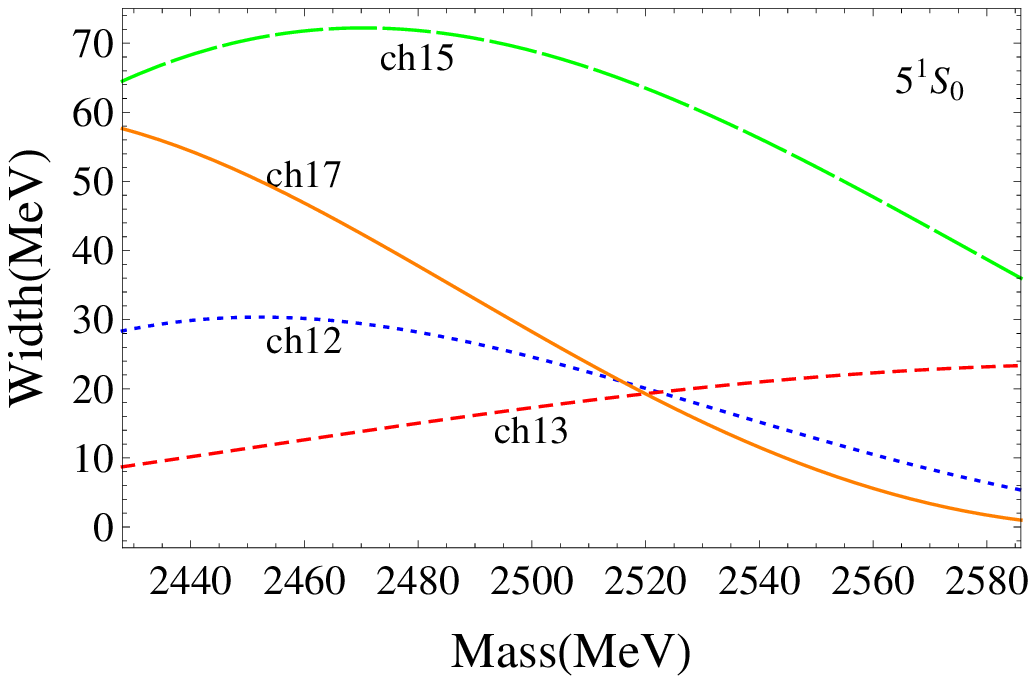}
\includegraphics[scale=0.6]{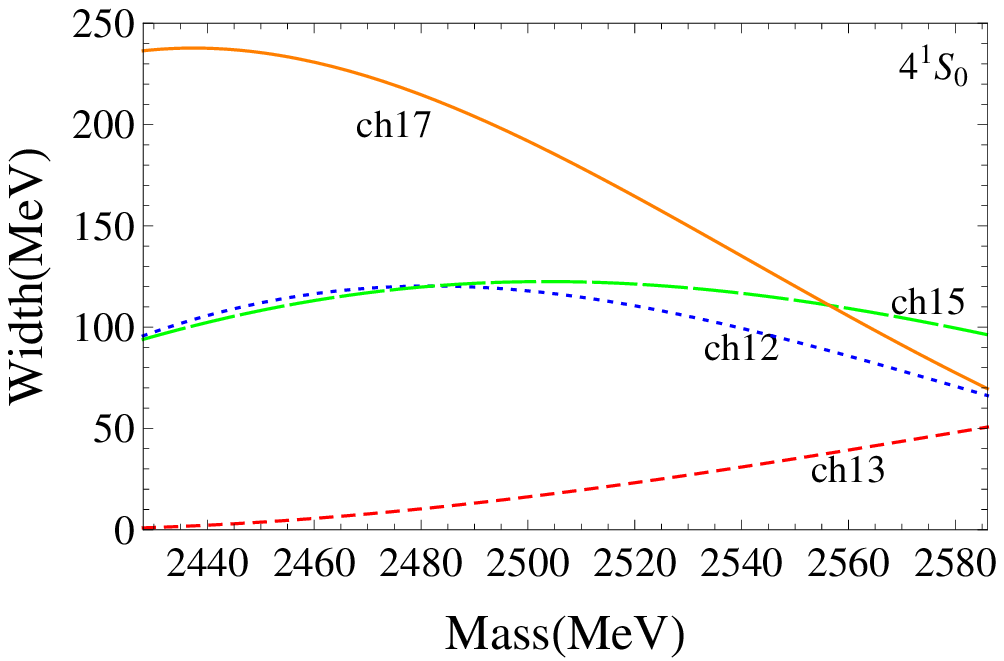}
\includegraphics[scale=0.6]{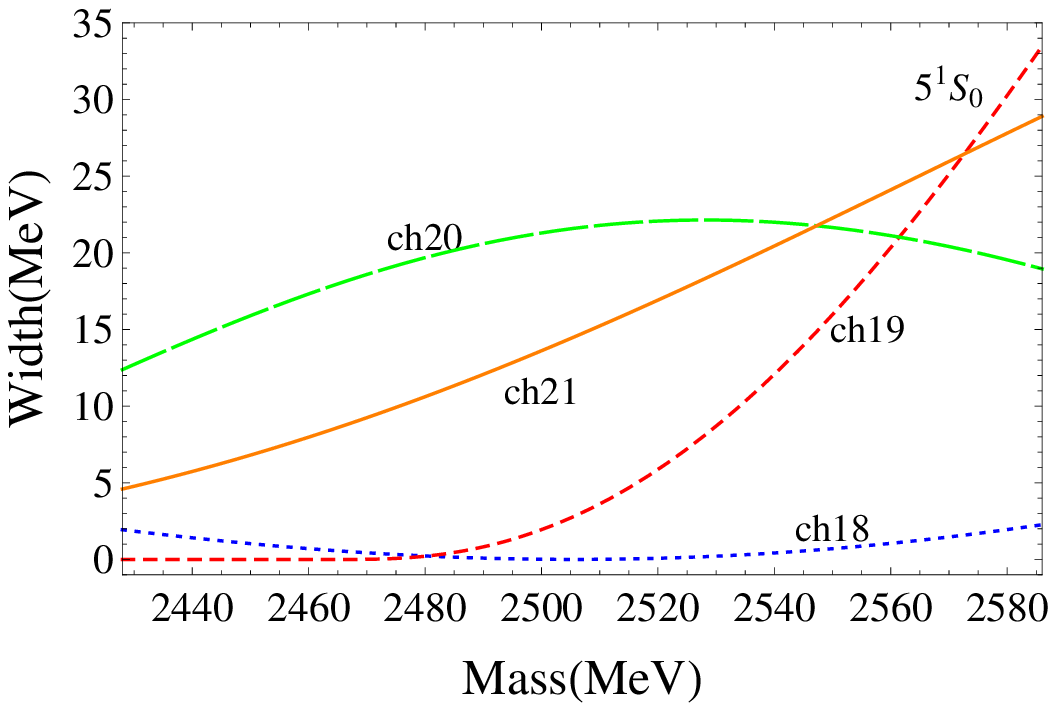}
\includegraphics[scale=0.6]{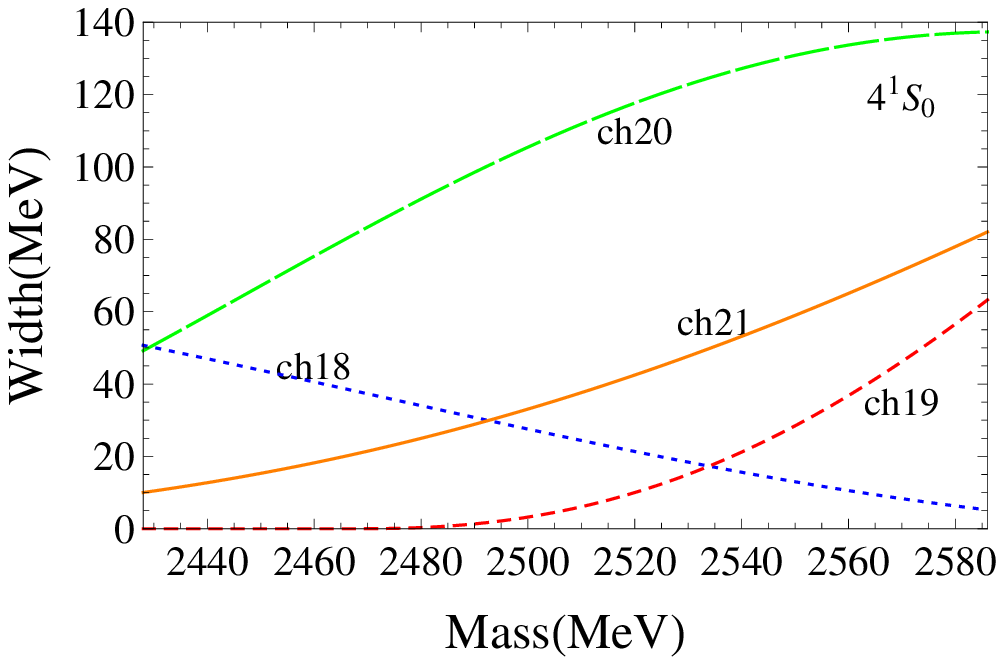}
\vspace{0.0cm}\caption{The dependence of the partial widths of the $5^1S_0$ (left panels) and  $4^1S_0$ (right panels) $s\bar{s}$ states on the initial state mass in the $^3P_0$ model. The curves labeled "ch$i$" stand for the results of the $i$ decay channels of Table~\ref{tab:decay}. }
\label{fig:pwidth}
\end{figure*}

 The parameters used in the $^3P_0$ model involve the $q\bar{q}$  pair production strength parameter $\gamma$, the SHO wave function scale parameter $\beta$, and the masses of the constituent quark. In this work, we choose to follow the Refs.~\cite{Li:2008et,a10,Li:2009rka} and take $\gamma=8.77$, $\beta_A=\beta_B=\beta_C=\beta=400$ MeV, $m_u=m_d=330$ MeV, and $m_s=550$ MeV\footnote{Our value of $\gamma$ is higher than that used by Ref.~\cite{a10} (0.505) by a factor of $\sqrt{96\pi}$, due to different filed conventions, constant factors in the transition operator $T$, etc. The calculated results of the widths are, of course, affected.}. The meson masses used in this work are
 $M_\eta=547.86$ MeV, $M_{\eta^\prime}=957.793$ MeV, $M_{\eta(1295)}=1294.4$ MeV, $M_{\eta(1475)}=1476$ MeV, $M_{f_0(980)}=990$ MeV, $M_{f_0(1710)}=1723$ MeV, $M_{\phi(1020)}=1019.46$ MeV, $M_{f_2^\prime(1525)}=1525$ MeV, $M_K=493.68$ MeV, $M_{K_0^*(1430)}=1425$ MeV, $M_{K_0^*(1950)}=1945$ MeV, $M_{K^*}=891.66$ MeV, $M_{K_1(1270)}=1272$ MeV,  $M_{K_1(1400)}=1403$ MeV, $M_{K(1460)}=1460$ MeV, $M_{K(1830)}=1830$ MeV, $M_{K^*(1410)}=1414$ MeV, $M_{K^*(1680)}=1717$ MeV, $M_{K^*_2(1430)}=1425.6$ MeV, $M_{K^*_2(1980)}=1973$ MeV, $M_{K^*_3(1780)}=1776$ MeV~\cite{Agashe:2014kda}. The meson flavor wave functions follow the conventions of Ref.~\cite{a9}.

With the above inputs, the decay widths of the $X(2500)$ as the $4^1S_0$ and $5^1S_0$ $s\bar{s}$ states are listed in Table~\ref{tab:decay}. The predicted total width of the $X(2500)$ as the $5^1S_0$ $s\bar{s}$ state
is 271.1 MeV, in agreement with the experiment data $\Gamma_{X(2500)}=230_{-35-33}^{+64+56}$ MeV within errors. If the $X(2500)$ is the $4^1S_0$ $s\bar{s}$ state, its total width is predicted to be about 894.5 MeV, much larger than the experimental data. The dependence of the predicted decay widths of the $X(2500)$ as the $4^1S_0$ and $5^1S_0$ $s\bar{s}$ on the initial state mass is shown in Fig.~\ref{fig:twidth}. As shown in Fig.~\ref{fig:twidth}, when the initial state mass varies from 2428 to 2580 MeV, the total width of the $5^1S_0$ $s\bar{s}$ state varies from 194 to 280 MeV, lying in the width range of the $X(2500)$, while the total width of the $4^1S_0$ $s\bar{s}$ state varies from 740 to 932 MeV, far more than the $X(2500)$ width. Therefore, it is difficult to explain the $X(2500)$ as the $4^1S_0$ $s\bar{s}$ state, but the assignment of the $X(2500)$ as the $5^1S_0$ $s\bar{s}$ state appears reasonable.

The smaller total decay width for the $5^1S_0$  assignment compared with the $4^1S_0$ assignment can be understood via the node structure of the wave function of the initial state. The nodes of the $5^1S_0$  state is more than that of the $4^1S_0$ state, and the overlap of the $5^1S_0$ state with the low-lying final states is smaller, which results in the smaller amplitude of the decay process. Hence, with the same phase space, the total decay width of the $4^1S_0$  state is much larger than that of the $5^1S_0$  state

The dependence of the partial strong decay widths of the $X(2500)$ as both $4^1S_0$ and $5^1S_0$ $s\bar{s}$ on the initial state mass is also presented in Fig.~\ref{fig:pwidth}, where the results for the $5^1S_0$ and $4^1S_0$ $s\bar{s}$ assignments are plotted in the left and right panels, respectively, and the curves labeled "ch$i$" stand for the results for the $i$ decay channels of Table~\ref{tab:decay}.  For both $5^1S_0$ and $4^1S_0$ $s\bar{s}$ states, the partial widths of the $\eta(1475) f_0(980)$, $KK_0^*(1950)$ and $K^*K^*(1400)$ channels are sensitive to the initial state mass.

The new pseudoscalar state $X(2500)$ has been observed in the decay $J/\psi\rightarrow \gamma X\rightarrow \gamma\phi\phi$~\cite{Ablikim:2016hlu}, however, the branching ratio of the $\phi\phi$ channel is predicted to be very small. So, in order to confirm or refute the possibility of the $X(2500)$ being the $5^1S_0$ $s\bar{s}$ state, the further confirmation of this small branching ratio is strongly called for.

Also, since only the $\phi\phi$ decay mode of the $X(2500)$ has been observed, an important test of this interpretation of the $X(2500)$ as $5^1S_0$ $s\bar{s}$ would be the observation of some of these other decay modes with large branching rations such as $KK(1830)$, $K^*K^*(1410)$, $K^*K(1460)$, $KK_0^*(1950)$, $K^*K_2^*(1430)$, $\eta(1475)f_0(980)$, and $K^*K_1(1400)$.

\section{SUMMARY AND CONCLUSION}
\label{sec:summary}

We have calculated the strong decay of the $X(2500)$ state with the assignments of $4^1S_0$ and $5^1S_0$ $s\bar{s}$ in the $^3P_0$ model. The predicted total width for the $4^1S_0$ $s\bar{s}$ is far from the the observed width of the $X(2500)$, while the one for the $5^1S_0$ $s\bar{s}$ is in good agreement with the experimental data. The mass of the $5^1S_0$ $s\bar{s}$
quantitatively estimated by Regge phenomenology is about 2.5 MeV~\cite{mesontrajectory,Masjuan:2012gc}, which is also consistent with the observed mass of the $X(2500)$.
Therefore, The available experimental evidence for the $X(2500)$ is in favor of the $5^1S_0$ interpretation.  To test this assignment, the further confirmation of the small branching ratio of $\phi\phi$ channel and the further information of other decay modes such as $KK(1830)$, $K^*K^*(1410)$, $K^*K(1460)$, $KK_0^*(1950)$, $K^*K_2^*(1430)$, $\eta(1475)f_0(980)$, and $K^*K_1(1400)$  are needed.

In addition, the accurate mass spectra of the high radial excited meson and strong decay properties of the pseudoscalar glueball can improve our understanding the $X(2500)$.

\section{Acknowledgements}

This work is partly supported by the National Natural Science Foundation of China under Grant No.11505158, the China Postdoctoral Science Foundation under Grant No.2015M582197, the Postdoctoral Research Sponsorship in Henan Province under Grant No.2015023,  and the Startup Research Fund of Zhengzhou University (Grants No. 1511317001 and No. 1511317002).

\end{document}